# Study on modeling 40t ultra high power (UHP) electric furnace

Yun Chol Guk, Ryang Gyong Il*

**Kim Il Sung** University, Pyongyang, DPR of Korea

Email Address: ryongnam26@yahoo.com

**Abstract**: In this paper we proposed a modeling method of the electrode lift controlling plant of an ultra high power electric furnace and verified the accuracy of the proposed model by simulation and field test.

## 1. Introduction

UHPs have become one of main tools for the steel making process since they have high productivity, low cost and high quality of products.

In this paper we make the mathematical model of the electrode lift controlling plant of 40t UHP that is the primary problem in developing the electrode lift control system.

In the previous work [1], they made the mathematical model of UHP, but they didn't consider the dynamics of UHP according to the melting stages.

In the previous work [2,3], they made the arc current prediction model using artificial intelligence technology and chaos theory, but they didn't consider the dynamics of UHP according to the melting stages.

The electrode lift controlling system of UHP regulates the driving voltage of the electrode lift equipment to sustain the electric arc length, so stabilizes the arc current and arc voltage.

The electrode lift controlling plant of the UHP consists of an electrode lift equipment and an electrode furnace (arc-length, impedance)(see figure 1 ).
Therefore the modeling of the electrode lift controlling plant of UHP consists of the modeling of the electrode lift equipment and the modeling of arc furnace (arc length- arc current, impedance).

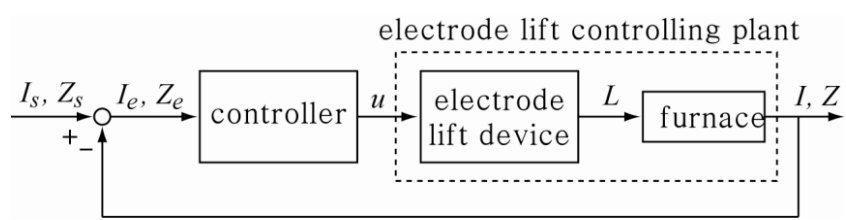

Figure 1. The electrode lift controlling plant of UHP.

## 2. Mathematical modeling the electrode lifting equipment.

The electrode lift equipment sustains the arc length to stabilize arc current, arc voltage and arc power. In this paper we model the electronic servo valve-hydraulics motoring electrode lift equipment (figure 2).

The electronic servo valve-hydraulics motoring electrode lift equipment regulates the valve position according to the driving voltage. The valve position regulates the oil to enter into and drain out the hydraulics cylinder, thus the electrode is lifted. When the oil amount increases, the electrode lift and when the oil amount decreases, the electrode down.

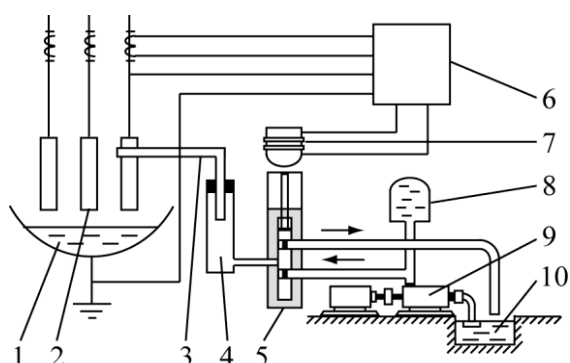

Figure2. Electronic servo valve-hydraulics driving electrode lifting equipment
(1－material, 2－electrode, 3－electrode lift equipment, 4－hydraulics cylinder, 5－electronic valve, 6－electrode lift controller, 7－moving magnet, 8－pressure tin, 9－hydraulics pump, 10－oil tank)

The input of the electrode lift equipment is the driving voltage u operates on the electronic valve and the output is the arc length. Table 1 shows the physical variables used in modeling of the electrode lift equipment.

Table 1. The physical variables used in the mathematical modeling of the electrode lift equipment

| № | symbol | meaning | unit | note |
|---|---|---|---|---|
| 1 | $\rho$ | Density of oil | $Kg/m^3$ | 850 |
| 2 | $g$ | gravity | $N/Kg$ | 9.8 |
| 3 | $m$ | Mass of the electrode | $Kg$ | 7 850 |
| 4 | $L$ | Arc length | $mm$ | |
| 5 | $u$ | driving voltage of the electronic valve | $V$ | |
| 6 | $Q$ | Flow | $m^3/s$ | |
| 7 | $P_1$ | Pressure lifting the piston | $Pa$ | |
| 8 | $P_2$ | Pressure press down by the mass of electrode | $Pa$ | |
| 9 | $F$ | Power acting on the piston | $N$ | |
| 10 | $A$ | Section area of the piston | $m^2$ | $1.54 \times 10^{-2}$ |
| 11 | $c$ | Viscosity | $N \cdot s/m$ | 0.05 |

The flow incoming into the pipe, Q is represented as a function the driving voltage u and the pressure difference of the cylinder, $P_c = P_1 - P_2$.

$$Q = f(u, P_c) \qquad (1)$$

We linearize equation (1) around $Q_0 = f(u_0, P_{c0})$ using Taylor series to obtain equation (2).

$$Q - Q_0 = \left.\frac{\partial Q}{\partial u}\right|_{u=u_0}(u - u_0) + \left.\frac{\partial Q}{\partial P_c}\right|_{P_c=P_{c0}}(P_c - P_{c0}) = k_1 \Delta u + k_2 \Delta P_c \qquad (2)$$

Where $k_1$ is the flow amplitude [$m^3/s \cdot V$] and $k_2$ is flow-pressure coefficient [$m^5/s \cdot N$]. We assume $Q_0 = 0$, $u_0 = 0$, $p_{c0} = 0$.

$$Q = k_1 u + k_2 P_c \qquad (3)$$

In other hand flow $Q$ is calculated as follows.

$$Q = A \frac{dL}{dt} = A \dot{L} \qquad (4)$$

The power lifting the piston is calculated as follows.

$$F = AP_1 = A(P_c + P_2) = AP_c + mg \qquad (5)$$

And As the power lifting the piston stands the inertia of the load, gravity, viscosity, following equation is formed.

$$AP_c + mg = m\ddot{L} + c\dot{L} + mg \qquad (6)$$

From equation (3) and (4)

$$P_c = \frac{1}{k_2}(Q - k_1 u) = \frac{1}{k_2}(A\dot{L} - k_1 u) \tag{7}$$

From equation (6) and (7)

$$\frac{A}{k_2}(A\dot{L} - k_1 u) + mg = m\ddot{L} + c\dot{L} + mg \tag{8}$$

To arrange the above equation,

$$m\ddot{L} + \left(c - \frac{A^2}{k_2}\right)\dot{L} = -\frac{k_1 A}{k_2} u \tag{9}$$

To perform Laplace's transform,

$$\frac{L(s)}{U(s)} = \frac{-\dfrac{k_1 A}{k_2}}{s\left(ms + c - \dfrac{A^2}{k_2}\right)} \tag{10}$$

To ignore the friction, let $c = 0$

$$\frac{L(s)}{U(s)} = \frac{-\dfrac{k_1}{k_2}A}{s\left(ms - \dfrac{A^2}{k_2}\right)} = \frac{K}{s(Ts + 1)} \tag{11}$$

Where $K = \dfrac{k_1}{A}$ and $T = -\dfrac{mk_2}{A^2}$ is the Amplitude and time const of the electrode lift equipment respectively.

Therefore the electrode lift equipment is considered a integral inertia part as equation (11).

And from equation (11) the transform function between the driving voltage and the electrode lift velocity is as equation (12), where the amplitude of the electrode lift equipment K and time const T may be determined by measuring the lift velocity of the electrode lift equipment when the driving voltage is +1V as unit step signal.

$$\frac{v(s)}{U(s)} = \frac{K}{Ts + 1} \tag{12}$$

At that time the amplitude $K$ is the normal value and the time const $T$ is the period that the response is 63.2% of the steady state. The parameters that are determined through the electrode lift experiment are as follows.

$$K = 15\ mm/s \cdot V,\ T = 0.1s$$

Therefore the transform function of the electrode lift equipment is as follows.

$$\frac{L(s)}{U(s)} = \frac{15}{s(0.1s+1)} \qquad (13)$$

## 3. Mathematical modeling a 40t arc furnace.

In an arc furnace, the arc current and the impedance are determined by the arc length. We obtain the relationship of the arc length L and the arc current I and the circuit impedance $Z$.

Table 2 show the physical terms used in modeling of the arc furnace.

Table 2. The physical terms used in modeling of the arc furnace.

| № | symbol | meaning | unit | note |
|---|---|---|---|---|
| 1 | $U_1$ | Primary line voltage | V | |
| 2 | $E_2'$ | Secondary phase voltage calculated in the primary side. | V | $E_2' = U_1 / \sqrt{3} k_T$ |
| 3 | $E_2$ | Secondary phase voltage | V | |
| 4 | $k_T$ | Ratio of the furnace transformer | | |
| 5 | $I$ | Arc current | A | |
| 6 | $U_a$ | Arc voltage | V | |
| 7 | $L$ | Arc length | mm | |
| 8 | $\beta$ | Voltage per unit arc length | V/mm | Melting stage: decrease from 12, oxidization stage:3.7 Revive stage:1.2 |
| 9 | $\alpha$ | Voltage between the anode and cathode | V | $\alpha = 9$ |
| 10 | $X$ | Impedance sum of primary circuit and the secondary circuit. | Ω | $X = (X_r + X_T)/k_T^2 + X_2$ |
| 11 | $X_r$ | Induction resister of the reactor | Ω | Dependent to tap of the reactor |
| 12 | $X_T$ | Induction resister of the furnace transformer | Ω | Dependent to tap of the transformer |
| 13 | $X_2$ | Whole reactance of the secondary circuit. | Ω | $3.0 \times 10^{-3}$ |
| 14 | $R_2$ | Whole resister of the secondary circuit. | Ω | $0.507 \times 10^{-3}$ |
| 15 | $Z$ | Impedance of the secondary circuit | Ω | $Z = \sqrt{(R_a + R_2)^2 + X^2}$ |

The circuit of 3 phase furnace is figure 3.

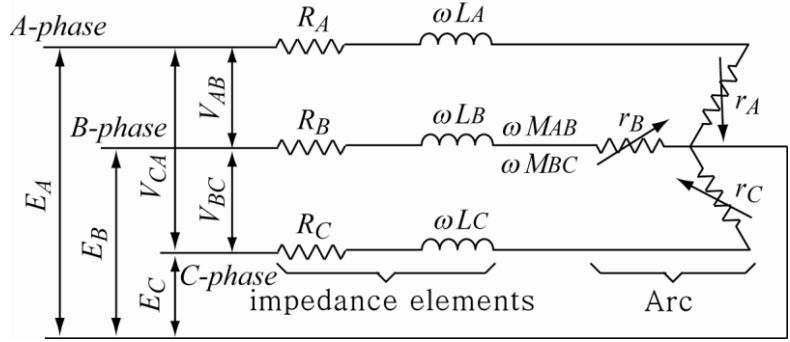

Figure 3. The circuit of 3 phase furnace

A 3-phase arc furnace is a 3-phase unbalanced circuit and the position of the neutral point of 3-phase circuit is varied. The single circuit is figure 4.

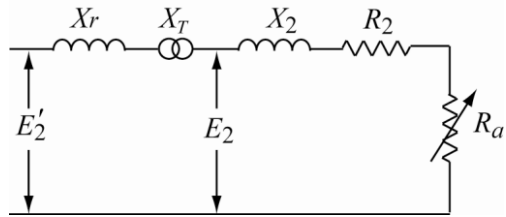

Figure 4. The single circuit of UHP

The vector diagram of the single circuit is figure 5.

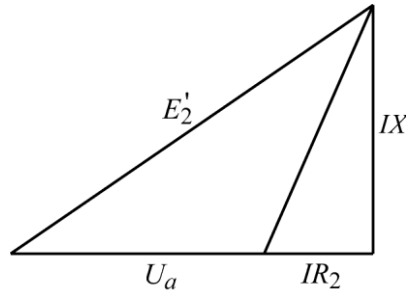

Figure 5. voltage vector circuit of the single circuit

$$( X = ( X_r + X_T ) / k_T^2 + X_2 )$$

From figure 6 we obtain equation (14),

$$E_2'^2 = (U_a + IR_2)^2 + (IX)^2 \tag{14}$$

Where the secondary phase voltage calculated in the primary side is $E_2' = U_1 / \sqrt{3} k_T$.

From above equation we can the following equation.

$$I = \frac{-R_2 U_a + \sqrt{(X^2 + R_2^2) E_2'^2 - X^2 U_a^2}}{X^2 + R_2^2} \tag{15}$$

And the following equation holds between the arc voltage and arc current,

$$U_a = \beta L + \alpha \tag{16}$$

Where $\alpha$ is the voltage between the anode and the cathode, in normal when we use carbon electrode, it is about $9V$ and $\beta$ is the voltage per unit arc length.

From equation (15) and (16), the following nonlinear equation between arc length and arc current holds.

$$I = f_1(L) = \frac{-R_2(\beta L + \alpha) + \sqrt{(X^2 + R_2^2)E_2'^2 - X^2(\beta L + \alpha)^2}}{X^2 + R_2^2} \quad (17)$$

And the following equation between impedance and arc current holds.

$$Z = \frac{E_2}{I} \quad (18)$$

Therefore the following nonlinear relationship between arc length and impedance holds.

$$Z = f_2(L) = \frac{(X^2 + R_2^2)E_2}{-R_2(\beta L + \alpha) + \sqrt{(X^2 + R_2^2)E_2'^2 - X^2(\beta L + \alpha)}} \quad (19)$$

As a result we can represent the electrode lifting control system of UHP as figure 6.

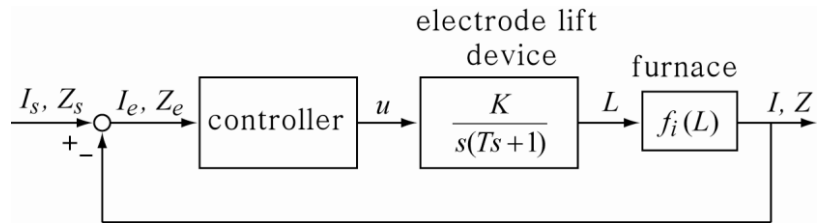

Figure 6. The electrode lifting control system of UHP

## 4. Validation

We verify the validity of the proposed model through simulation and the field test.

At first by using MATLAB Simulink simulation as figure 7, we obtain the unit step response of the electrode lift controlling plant that is consisted of the electrode lift equipment and the arc furnace, we compare it with the field test data (arc current curve obtained when we operate +1V of driving voltage on the electrode lift equipment).

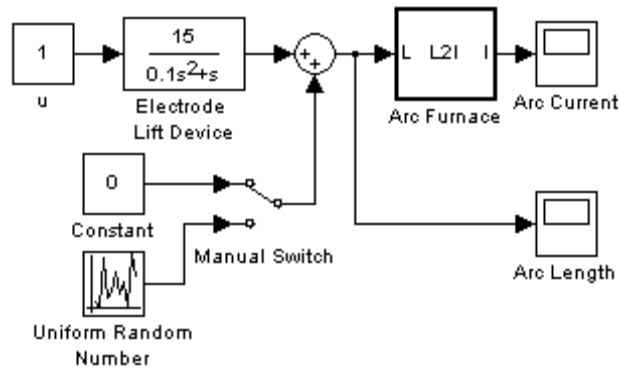

Figure 7. Simulink diagram for the model validation.

Figure 8 shows the curves of model response and real plant response together.

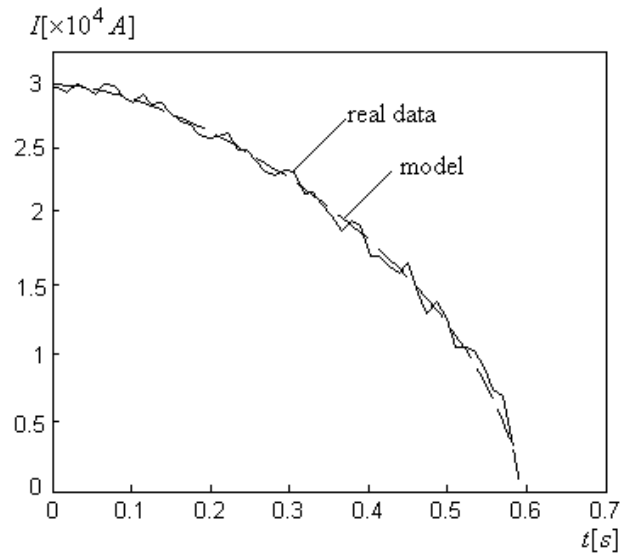

Figure 8. Model validation ($\beta = 12$, $tap = 7$)

The fitness of the model is 95.57%. This model can be used in the design of the electrode lift controller.

## 5. Conclusion

In this paper, we have developed a mathematical model of 40t UHP electrode lift equipment. The electronic servo valve-hydraulic motor electrodes lift equipment I has been approximated as first order inertia and integral element and the nonlinear relations of arc length and arc current, impedance have been developed, and then we validated it through the model validation.